\documentclass[aps,prc,onecolumn,preprint,showpacs]{revtex4} 

\usepackage{amsmath} 
\usepackage{amssymb} 
\usepackage{graphicx} 
\usepackage{graphics}
\usepackage{epsfig}
\usepackage{multirow}
\usepackage{color}
\usepackage{slashed}
\newcommand\nn{\nonumber}
\newcommand\ba{\begin{eqnarray}}
\newcommand\ea{\end{eqnarray}}
\newcommand\be{\begin{equation}}
\newcommand\ee{\end{equation}}
\newcommand\bi{\begin{itemize}}
\newcommand\ei{\end{itemize}}
 
\begin{document}
\title{Antiproton--proton annihilation into light neutral meson pairs within an effective meson theory}

\author{Ying~Wang\footnote{Chinese CSC Scholar}} 
\affiliation{Institut de~Physique~Nucl\'eaire,~Universit\'e~Paris-Saclay,~91405~Orsay,~France}
\author{Yury~M.~Bystritskiy \footnote{E-mail: bystr@theor.jinr.ru}} 
\affiliation{Joint~Institute~for~Nuclear Research,~141980~Dubna,~Russia}

\author{Azad~I.~Ahmadov\footnote{E-mail: ahmadov@theor.jinr.ru}}
\altaffiliation{Institute~of~Physics,\- Azerbaijan~National~Academy~of~Sciences,~Baku,~Azerbaijan}
\affiliation{Joint~Institute~for~Nuclear Research,~141980~Dubna,~Russia}

\author{Egle~Tomasi-Gustafsson \footnote{Corrresponding author: egle.tomasi@cea.fr}} 
\affiliation{IRFU,~CEA,~Universit\'e~Paris-Saclay,~91191~Gif-sur-Yvette,~France}

\date{\today}
\begin{abstract}
Antiproton--proton annihilation into light neutral mesons  in the few GeV energy domain is investigated in view of a global description of the existing data and predictions for future experiments at PANDA, FAIR. An effective meson model earlier developed, with mesonic and baryonic degrees of freedom in $s$, $t$, and $u$ channels, is applied here to $\pi^0\pi^0$ production. Form factors with logarithmic $s$ and $t(u)$ dependencies are applied.  A fair agreement with the existing angular distributions is obtained. Applying SU(3) symmetry, it is straightforward to recover the angular distributions for $ \pi^0\eta$, and $\eta\eta$ production in the same energy range. A good agreement is generally obtained with all existing data. 
\end{abstract}
\maketitle
\section{Introduction}
In a previous paper \cite{Wang:2015ybw} we proposed an effective Lagrangian model with meson and baryon exchanges in $s$, 
$t$, and $u$ channels ($s$, $t$ and $u$ are standard kinematical Mandelstam variables) to describe  the exclusive annihilation reaction of antiproton-proton annihilation into charged pion and kaon pairs in the energy domain  ($2.25 (1.5) \le \sqrt{s} (p_L ) \le 5.47(15)$ GeV (GeV/c)) where $\sqrt{s} (p_L )$ is the total energy(the beam momentum) in the Laboratory (Lab) frame.  This is the domain relevant to the antiProton ANnihilation at DArmstadt (PANDA) experiment at the GSI Facility for Antiprotons and Ion Research (FAIR) \cite{Lutz:2009ff}. Data in this energy range are scarce, poorly constraining the models. To validate our approach we considered also pion-proton elastic scattering data through crossing symmetry. 

A large amount of data on light meson production is expected in near future. In the PANDA energy range, exclusive charged and neutral pion pair productions in $\bar p p$ collisions  bring information on the non perturbative structure of the proton and on the hadronization mechanisms.  In the low energy region, particularly studied  at the Low Energy Antiproton Ring (LEAR) at CERN,  the angular distributions show a series of oscillations, typically reproduced by Legendre polynomials, describing contributions of higher excitations $L$-waves. Potential models  \cite{Loiseau:1992xe} successfully describe the cross sections and polarization observables but their extension to higher energies becomes too involved.  Increasing the energy, the angular distributions loose progressively the oscillating behavior. Above $\sqrt{s}$=2 GeV two body  processes become mostly peripheral and the angular distributions are peaked forward or backward, corresponding to small values of $t$ or $u$, respectively.  The cross section for pions emitted at 
$\cos\theta=0$ ($\theta$ is the emission angle in the center of mass system (CMS)) shows a scaling behavior as a power of $s$, near to $s^{-8}$ , as predicted by QCD quark counting rules, whereas for forward and backward scattering, the $s$ dependence is near to exponential. A brief review of relevant data and models can be found in our previous work Ref. \cite{Wang:2015ybw}. 

We extend here the model developed in Ref. \cite{Wang:2015ybw} to $\bar p p$ annihilation into neutral meson pairs. As in Ref. \cite{Wang:2015ybw}, $t$ and $u$ exchanges of nucleon and $\Delta$ are considered. First order Born diagrams are calculated and form factors are added.  Instead than monopole, dipole or exponential form factors, $i.e.$, the functional forms that can be found in the literature, we propose $s$ and $t$ dependent logarithmic form factors, after being convinced that the Regge regime is not yet applicable in the considered energy region. Compared to charge pion production, the necessary modifications are the symmetrization of the final state for identical mesons and the nature of the exchanged meson in $s-$channel. Since $\rho$-exchange is forbidden by $G$-parity, the lighter mesons that can be exchanged are the scalar $f_0$ and $f_2$ mesons, with mass and width as:  \cite{Agashe:2014kda}:
\ba
& f_0(500) I^G(J^{PC})=0^+(0^{++})& m_{f_{0}}=(400-550)\  MeV, Ê\ \Gamma_{f_0} =(400-700) \ MeV \nn\\
& f_0(980) I^G(J^{PC})=0^+(0^{++})& m_{f_0}=(990\pm 20)\  MeV, Ê\ \Gamma_{f_0} =(40-100) \ MeV \nn\\
& f_2(1270) I^G(J^{PC})=0^+(2^{++})& m_{f_2}=(1275,5\pm 0.8)\  MeV, Ê\ \Gamma_{f_2} =(186.7\pm 2.5) \ MeV 
\ea
Pion emission around $\cos\theta=0$ is driven by $s$-channel exchange. We limit  our considerations to $s$-channel $f_0$ and $f_2$-meson exchange. In case of $f_0$ we take 'an effective $f_0$' with  mass $M=600$ MeV and width $\Gamma = 700$ MeV. In principle, other higher mass resonances that decay into $\pi^0\pi^0$ may be considered. However they are suppressed outside the resonance peak due to the Breit-Wigner representation of the corresponding amplitudes. An additional suppression of radial excitations of these mesons is expected because their spatial density is less compact, making  less probable the formation of a pion pair. Exclusive pion pairs are formed with the largest probability  when the two $q\bar q$ pairs emerge from the vacuum in a physical space-time region with small dimension.

We compare our calculation to the data on neutral pion (and other neutral meson) production,  published by the FermiLab E760 collaboration in the energy range ($2.911\le \sqrt{s} \le 4.274$) GeV \cite{Armstrong:1997gv}.  The primary aim of that work was to study heavy meson resonances that couple to $\bar N N$, as charmonium. Moreover the study of the $s$-dependence in terms of power laws showed that an approximate scaling is reached, but with lower exponent than predicted.  The measured angular distributions are limited to a central angular range, $|\cos\theta| \le$0.66. At the lowest energies, the $\pi^0\pi^0$ angular distribution shows a bump at $|\cos\theta| =0$,  that gradually disappears from 2.9 to 3 GeV, and can be reproduced  including higher $L$-multipolarities, only. At our knowledge, at present, no calculation attempting to reproduce the whole set of data from Ref. \cite{Armstrong:1997gv}  exists in the literature.  

Our aim is to build a reliable and coherent model that reproduces  the basic features of neutral meson production in the energy range that will be investigated by the future experiment PANDA at FAIR. With the help of SU(3) symmetry, we apply our model to other neutral channels $\eta\eta$  and $\pi^0\eta$, where data are present. The model should have minimal ingredients, and analytical expressions, convenient to be included in the PANDARoot Monte Carlo simulation program.

\section{Formalism}
\subsection{Kinematics and cross section}

We consider the annihilation reaction:
\be
\bar p(p_1)+p(p_2)\to \pi^0(k_1)  + \pi^0(k_2) ,
\label{Eq:eq1} 
\ee
in CMS. The  notation of four momenta is shown in the parenthesis. The following notations are used: $q_t=-p_1+k_1$, $q_t^2=t$, $q_u= -p_1+k_2 $, $q_u^2=u$, and $q_s=p_1+p_2 $, $q_s^2=s$, $s+t+u=2M_N^2+2m_\pi^2$, $M_N$($m_\pi$) is the nucleon(pion) mass (for reactions (\ref{Eq:eq2},\ref{Eq:eq3}) the corresponding mass should be substituted). The useful scalar product between four vectors are explicitly written as:
\ba
&2p_1 k_2=2k_1p_2=M_N^2+m_\pi^2-u,\ 
&2p_1 k_1=2k_2p_2=M_N^2+m_\pi^2-t,\nn\\
&2p_1 p_2=s -2M_N^2,\ 
&2k_1k_2=s -2m_\pi^2, \nn\\
&p_1^2=p_2^2=M_N^2=E^2-|\vec p|^2, \ 
&k_1^2=k_2^2=m_\pi^2=\varepsilon^2-|\vec k|^2. \ 
\ea
In particular ,the final particles mass-shell conditions fixes the energies $E_{1,2}$,  the velocity  $\beta_{1,2}$
and the modulus of the momentum $\vec k$  of the final particles (where "1" refers to the detected particle, and "2" to the partner):
\be
    E_{1,2} = \frac{s+M_{1,2}^2-M_{2,1}^2}{2\sqrt{s}}, \ 
    \beta_{1,2} = \frac{\lambda^{1/2}(s,M_{1,2}^2,M_{2,1}^2)}{s+M_{1,2}^2-M_{2,1}^2}, \ 
     |\vec k | = \frac{1}{2\sqrt{s}}\,\lambda^{1/2}(s,M_1^2,M_2^2).
    \label{eq.eqk}
    \ee
where $\lambda(x,y,z)$ is the so called triangle function:
\be
    \lambda(x,y,z) = x^2 + y^2 + z^2 - 2\, x y - 2\, x z - 2\, y z . 
     \label{eq.eqe}
\ee

The general expression for the differential cross section in the CMS of reaction (\ref{Eq:eq1}) is:
\be
\displaystyle\frac {d\sigma}{d\Omega}= 
\displaystyle\frac{1}{2^8\pi^2 } \displaystyle\frac{1}{s}
\displaystyle\frac{\beta_{\pi} }{\beta_p }\overline{|{\cal M}|^{2}},\ \ 
\displaystyle\frac{d\sigma}{d\cos\theta}= 2E^2\beta_p\beta_{\pi} \displaystyle\frac{d\sigma}{dt}, 
\label{Eq:tcs}
\ee
where ${\cal M}$ is the amplitude of the process, $\beta_{p} $($\beta_{\pi}) $ is the velocity and $E(\varepsilon)$ is the energy of the proton(pion) in CMS. The phase volume can be transformed as 
$d\Omega \to 2\pi \ d\!\cos\theta$ due to the azimuthal symmetry of binary reactions.
The total cross section then reads as:
\be
\sigma=\int \frac{\overline{|{\cal M}|^2}}{64 \pi^2 s} \frac {|\vec p|}{|\vec k|}d\Omega , 
\label{eq:stot}
\ee
where $|\vec p|$ is the initial  momentum and $|\vec k|$ the momentum of the final detected particle  in CMS. In case of identical particles one should integrate only on half of the phase volume. $\overline{|{\cal M}|^2}$ is the squared matrix element of the process averaged over the spins of the initial particles.

\subsection{The reaction mechanism}
The formulas written above are model independent, $i.e.$, they hold for any reaction mechanism. In order to calculate 
${\cal M}$, one needs to specify a model for the reaction. In this work we consider the process (\ref{Eq:eq1})  within the formalism of effective meson Lagrangian. 

\begin{figure}[htp!]
\begin{center}
\includegraphics[width=14cm]{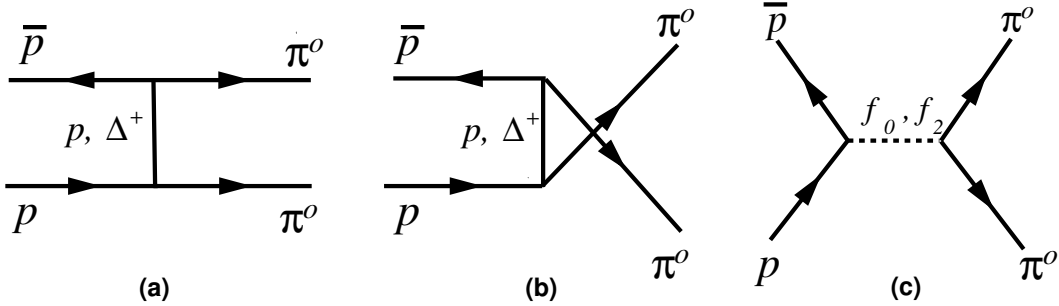}
\caption{Feynman diagrams for different exchanged particles for the reaction $\bar p+p \to \pi^0 +\pi^0$.}
\label{Fig:DiaAll}		
\end{center}
\end{figure}

The following contributions to the cross section for reaction (\ref{Eq:eq1}) are calculated as illustrated in Fig. (\ref{Fig:DiaAll}):
\begin{itemize}
\item  baryon exchange:
\begin{itemize}
\item  $t$-channel nucleon (neutron)  and $\Delta^+$ exchange, Fig. \ref{Fig:DiaAll}.a,
\item corresponding $u$-channel, crossed leg diagrams, Fig. \ref{Fig:DiaAll}.b,
\end{itemize}
\item $s$-channel  $f_0$, $f_2$  exchange,  Fig. \ref{Fig:DiaAll}c.
\end{itemize}

After the calculation of the coupling constant and matrix elements, the total 
matrix element squared averaged over the spin states of the initial particles, is obtained as the sum of the squared of matrix element for the individual contributions and the interferences among them. Identical particles in the final channel ($\pi^0\pi^0$ or $\eta\eta$ ) require to symmetrize the amplitudes. The matrix element squared, obtained from the coherent sum of the amplitudes, is:
\be
\overline{|{\cal M}|^{2}}=\displaystyle\frac{1}{\sqrt{2}}\overline{ |{\cal M}_p(t) + {\cal M}_{\Delta^+}(t) + {\cal M}_{f}(s) + {\cal M}_p(u) + {\cal M}_{\Delta^+}(u)  |^2}. 
\label{Eq:eqAll}		
\ee
Explicitly: 
\ba
|{\cal M}(\bar p p \to \pi^0\pi^0)|^{2}&=&
 |{\cal M}_{f_0}(s)|^2+|{\cal M}_{f_2}(s)|^2+\displaystyle\frac{1}{2}\Bigl \{ |{\cal M}_p(t)|^2 + |{\cal M}_{\Delta}(t)|^2 +
 \nn\\
&&
 |{\cal M}_p(u)|^2 + |{\cal M}_{\Delta}(u)|^2+
2Re[
{\cal M}_p(t)^*{\cal M}_p(u) + 
{\cal M}_p(t)^*{\cal M}_{\Delta}(t) +
\nn\\
&&{\cal M}_p(t)^*{\cal M}_{\Delta}(u) +
{\cal M}_p(u)^*{\cal M}_{\Delta}(t) +
{\cal M}_p(u)^*{\cal M}_{\Delta}(u) ] \Bigr \}+
 \nn\\
&&
\sqrt{2} Re[{\cal M}^*_{f_0}(s){\cal M}_{f_2}(s)+{\cal M}_p(t){\cal M}^*_{f_0}(s)+
{\cal M}_p(u){\cal M}^*_{f_0}(s) +
\nn\\
&&
{\cal M}_{\Delta}(t){\cal M}^*_{f_0}(s)+
 {\cal M}_{\Delta}(u){\cal M}^*_{f_0}(s)+{\cal M}^*_p(t){\cal M}_{f_2}(s)+
\nn\\
&&
{\cal M}^*_p(u){\cal M}_{f_2}(s) +{\cal M}^*_{\Delta}(t){\cal M}_{f_2}(s)
+ {\cal M}^*_{\Delta}(u){\cal M}_{f_2}(s) ].
\label{Eq:eqM2}
\ea
Taking into account the phase space and the flux, the expression for the total cross section is:
\be
\displaystyle\frac{d\sigma}{d\Omega}(\bar p p \to \pi^0\pi^0)=
\displaystyle\frac{1}{2^8\pi^2}
\displaystyle\frac{1}{s}
\displaystyle\frac{\beta_{\pi}}{\beta_p} |{\cal M}(\bar p p \to \pi^0\pi^0)|^{2},
\label{Eq:dsdol}
\ee
or 
\be
\displaystyle\frac{d\sigma}{d\cos\theta}(\bar p p \to \pi^0\pi^0)=
\displaystyle\frac{1}{2^7\pi}
\displaystyle\frac{1}{s}
\displaystyle\frac{\beta_{\pi}}{\beta_p} |{\cal M}(\bar p p \to \pi^0\pi^0)|^{2}.
\label{Eq:dsd}
\ee

For the explicit expressions of the $t-$ and $u$ channel $N$ and $\Delta$ amplitudes, in Eq. (\ref{Eq:eqM2}) we refer to the Appendix of Ref. \cite{Wang:2015ybw}. Coupling constants are fixed from the known decays of the particles when possible, otherwise values from effective potentials as \cite{PhysRevC.63.024001} are used. Masses and widths are taken from PDG \cite{Agashe:2014kda}.

\begin{figure}[htp!]
\mbox{\includegraphics[width=6.cm]{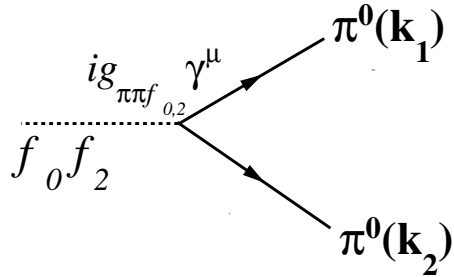}}
\caption{Diagram for $f_0$ and $f_2$ decays into a pion pair.}
\label{Fig:f02vertex}
\end{figure}

Let us consider the $f_0(500)$ also called $\sigma$ meson, the lowest isoscalar scalar particle, with spin zero and positive parity, and the next higher $L$ contributions, the $f_2(1270)$ with spin 2 and positive parity. Both decay dominantly  into two neutral pions (see Fig. \ref{Fig:f02vertex}).
\bi
\item  The $f_{0,2}$ propagators are taken as a Breit-Wigner function :  
\be
 \displaystyle\frac{1}{q_s^{2}-m_{f_{0,2}}^{2}+i \sqrt{q_s^2}\Gamma_{f_{0,2}}(q_s^2) }, 
 \ee
and the transferred momentum is $q_s=p_1+p_2=k_1+k_2$, $q_s^2=s$.
\item
 the  $f_0 \pi\pi$ vertex is $ -i g_{f_0  \pi\pi}$, being  $g_{f_0  \pi\pi}$ the constant for the decay $f_0 \to  \pi^0\pi^0$ (see Appendix A). The final expression for the width is: 
\be
 \Gamma_{f_{0}} =\displaystyle\frac{1}{16 m_{f_{0}}\pi} {g_{f_0\pi\pi}^2} \sqrt{1- \displaystyle\frac{4 m_{\pi}^2}{m_{f_{0}}^2} }, 
\label{Eq:phaseR2}   
\ee
where Taking the value: $\Gamma_{f_{0}}= 700 \pm150$ MeV  (in the range suggested by PDG \cite{Agashe:2014kda}), one finds  $g_{f_0\pi\pi}=4.08\pm 1.3$ GeV. 
\item  the  $f_0 NN$ vertex is $-ig_{f_0 N N }$   where $g_{f_0  NN}= 5 $ GeV  is the coupling constant  from Ref. \cite{PhysRevC.63.024001}.

\item The expression for the width of the decay $f_2\to \pi \pi$ is (see Appendix B3): 
\be
 \Gamma_{f_{2}} = \displaystyle\frac{g^2_{f_2\pi\pi}}{16m_{f_2}\pi} |{\cal M}(f_{2}\to \pi \pi)|^2\sqrt{1- \displaystyle\frac{4 m_{\pi}^2}{m_{f_{2}}^2} }.
\label{Eq:phaseR3}   
\ee
Taking the value: $\Gamma_{f_{2}}= (0.1867\pm 0.0025) $ GeV, one finds  $g_{f_2\pi\pi}= (19 \pm 0.26) $ GeV$^{-1}$. 

\item The vertex $f_2\to p p$ then is written as :
\be
(-i) g_{f_{2}pp}\gamma_\mu (p_1-p_2)_\nu \chi^{\mu\nu}.
 \label{Eq:gam}
 \ee
 where $g_{f_{2}pp}$ is considered as a fitting parameter and $\chi^{\mu\nu}$ is defined in Appendix B.
\ei
\section{Results}

The following procedure was applied, in order to reproduce the collected data basis. The data on neutral pion angular distributions from Ref. \cite{Armstrong:1997gv}  were first reproduced at best, with particular attention to the $s$ dependence of the cross section. The necessary number of parameters is very limited and we checked that the results are quite stable towards a change of the parameters in a reasonable interval.

The composite nature of the hadrons should be taken into account in the calculation of the observables. In order to find the best description of the data in a wide energy and angular ranges, different choices for form factors can be found in the literature: monopole, dipole, exponential etc. In Ref. \cite{Wang:2015ybw} a function of  logarithmic type turned out to reproduce at best  the measured angular and energy dependencies. The background of this choice is  a QCD derivation from  Refs.\cite{chernyak1977asymptotic,lepage1979exclusive} that relates the asymptotic behavior of form factors to the quark contents of the participating hadrons. It is also known that a logarithmic dependence of the $\bar p p$ cross section reproduces quite well the background for resonant processes \cite{Ablikim:2014jrz,PhysRevD.73.012005}. 

The logarithmic functional form  is:
\be
F^L _{N,\Delta}(x)=\displaystyle\frac {{\cal N}_{N,\Delta}\cdot M_0^4}{\left [(x-\Lambda_{N,\Delta}^2)\log\displaystyle\frac {(x-\Lambda_{N,\Delta}^2)}{\Lambda_{QCD}^2}\right ]^2},~ x=s,t,u,~M_0=3.86 ~ \mbox{GeV},~ \Lambda_{QCD}=0.3~ \mbox{GeV} ,
\label{eq:LogFF}
\ee
where  $M_0$ is a scale parameter, that has been inserted to conserve units, $ \Lambda_{QCD}$ is the QCD scale parameter. 
${\cal N}_{N,(\Delta)}=0.361 \pm 0.006 (0.041\pm 0.003)$ is a normalization constant. ${\Lambda}_{N,(\Delta)}= 2.25\pm 0.09 (1.05\pm 0.04)$ GeV is a ``slope" parameter which values were determined from a fit on the available data on charged pion production. A summary of parameters is listed in Table \ref{Table:pich} for nucleon and $\Delta$ exchange. 

\begin{figure}[htp!]
\mbox{\includegraphics[width=8.cm]{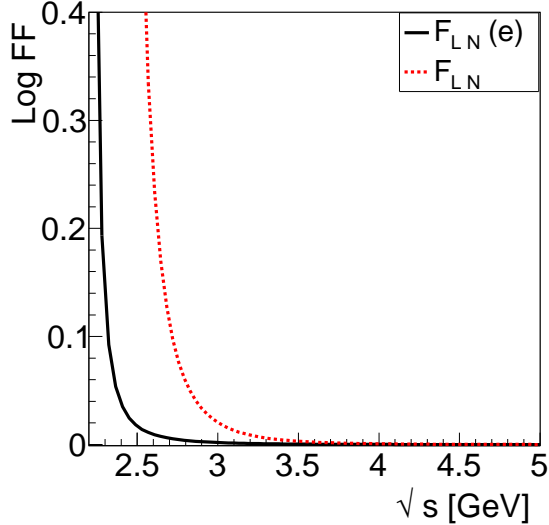}}
\caption{Energy dependence of the logarithmic form factors without (red, dashed line) and with (black, solid line) exponential correction.}
\label{Fig:FFSdep}
\end{figure}

For neutral pion pair production, the first attempt was to apply the same form factors and the same parameters as for the charged pion data for  $t(u)$ N and $\Delta$ exchanges from \cite{Wang:2015ybw},  the $s$ channel being calculated apart as physics requires the exchange of different mesons.  Similarly to charged meson production, first we apply the form factor $F^L_{N,\Delta}$ (Eq. \ref{eq:LogFF}) which depends on momentum transfer ($t$ or $u$) to take into account the composite nature of the particle in the interaction point. Second, we use the factor $F^L_{N,\Delta}(s)$ which effectively takes into account pre-Regge regime excitations of higher resonances in the intermediate state. This leads to an effective form factor as the product:
\be
 \widetilde{F}_{N,\Delta}(s,t)=F^L_{N,\Delta}(s)F^L_{N,\Delta}(t) ~(or \widetilde{F}_{N,\Delta}(s,u)=F^L_{N,\Delta}(s)F^L_{N,\Delta}(u)),
 \label{Eq:LogAng}
 \ee
containing the same set of parameters for the  $s$ and $t(u)$ dependencies, but different for $N$ and  $\Delta$ exchanges. The fit does not require independent parameters for $s$ and $t(u)$ dependencies. 
\begin{table}
\begin{tabular}{|c|c|}
\hline\hline
Parameter & Value \\
\hline
${\cal N}_{N}$ &  0.361        $\pm$  0.006 \\
${\cal N}_{\Delta}$ &0.041   $\pm$  0.003  \\
$\Lambda^2_{N}$ & (2.25     $\pm$    0.09) GeV$^2$\\
$\Lambda^2_{\Delta}$  &(1.05 $\pm$ 0.04 ) GeV$^2$ \\
\hline\hline
\end{tabular}
\caption{Summary of the parameters for the logarithmic form factors Eq. (\ref{Eq:LogAng}).}
\label{Table:pich}
\end{table}
The behavior of the total cross section for charged and neutral pion pair production is, however, very different. A possibility for recovering the $\pi^0\pi^0$ data is to modify the $s-$dependent part of the logarithmic form factors by adding an additional  energy dependence to the parameters:
\ba
{\cal N}(s)_{p,\Delta} &\to& {\cal N}(s)_{p,\Delta}-e^{\frac{p^{\cal N}_{p,\Delta}(s)}{\sqrt{s}}}, \nn\\
\Lambda(s)^2_{p,\Delta}& \to& \Lambda(s)^2_{p,\Delta}-e^{\frac{p^{\Lambda}_{p,\Delta}(s)}{\sqrt{s}}}.
\label{eq:s-exp}
\ea
In Fig. \ref{Fig:FFSdep} one can see the effect of the introduced $s$-dependence. The parameters converge at high energies, whereas for $\sqrt{s}\le 3.5$ GeV they deviate essentially, giving further reduction of the cross section. 
The $s$-independent parameters are fixed as in Table \ref{Table:fixed}.
\begin{table}
\begin{tabular}{|c|c|}
\hline\hline
Parameters & Value [GeV] \\
\hline
$p^{\cal N}_{p}(s)$ & -3.013 $\pm$ 0.210  \\
$p^{\cal N}_{\Delta}(s) $&  -5.959 $\pm$ 0.205\\
$p^{\Lambda}_{p}(s)$ & 4.047 $\pm$ 0.019 \\
$p^{\Lambda}_{\Delta}(s)$ &3.141 $\pm$ 0.002 \\
\hline\hline
\end{tabular}
\caption{Parameters for the $s-$dependent term of the logarithmic form factors for $\bar p p \to \pi^0\pi^0$ .}
\label{Table:fixed}
\end{table}

The form factor for the $f_0 NN$ vertex is taken of monopole form:
\be
{\cal F}_{f_0} (s) =\frac{F_{f_0}^2}{F_{f_0}^2+(m^2_{f_0}-s)},
\label{eq:f0}
\ee
with $F_{f_0}=1.17 \pm 0.051$ GeV. In addition, similarly to the charged pion calculation, the phase $\Phi_{f}= e^{i\pi \phi_f}$ is added for the exchanged meson in $s$-channel with $\phi_f$ equal to unity. 

%
\section{Comparison with existing data}

\begin{figure}[htp!]
\mbox{\includegraphics[width=16.cm]{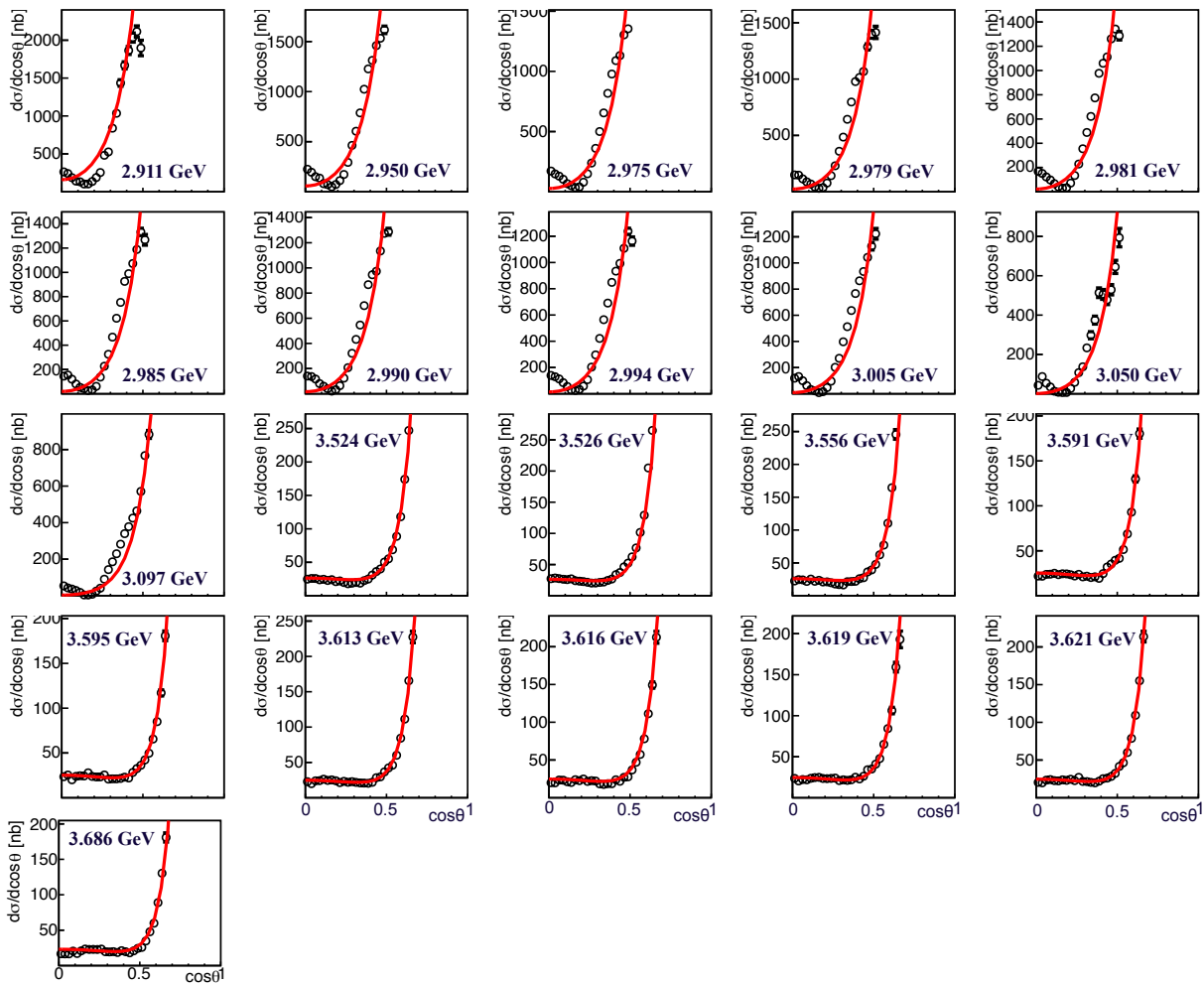}}
\caption{Angular distribution for the reaction $\bar{p} p \rightarrow \pi^0\pi^0$ in CMS in the energy range 2.911 GeV $\le \sqrt{s}\le$ 3.686 GeV. The data (open circles) are from Ref. \cite{Armstrong:1997gv}. The calculation is the solid, red line.}
\label{Fig:lowhigh}
\end{figure}

\begin{figure}[htp!]
\mbox{\includegraphics[width=8.cm]{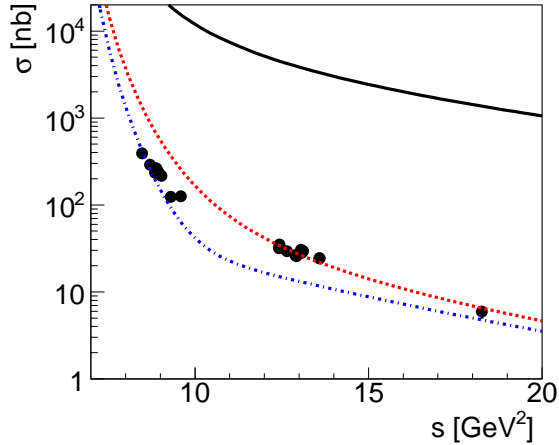}}
\caption{Integrated cross section for the reaction $\bar{p} p \rightarrow \pi^0\pi^0$. The data are obtained by the integration of the partial differential cross section in the available range: $0<\cos\theta<  0.48$ up to $\sqrt{s}=3.2$ GeV, and $0<\cos\theta<  0.66$ above $\sqrt{s}=3.6$ GeV,  Ref. \cite{Armstrong:1997gv}. The present calculation covering the range $0<\cos\theta<  0.48$ ( blue dash-dotted line)  and $0<\cos\theta<  0.66$ (red dashed line) is also shown. The integration in the whole angular range is shown as a black, solid line.}
\label{Fig:pi0total-s}
\end{figure}

\begin{figure}[htp!]
\mbox{\includegraphics[width=8.cm]{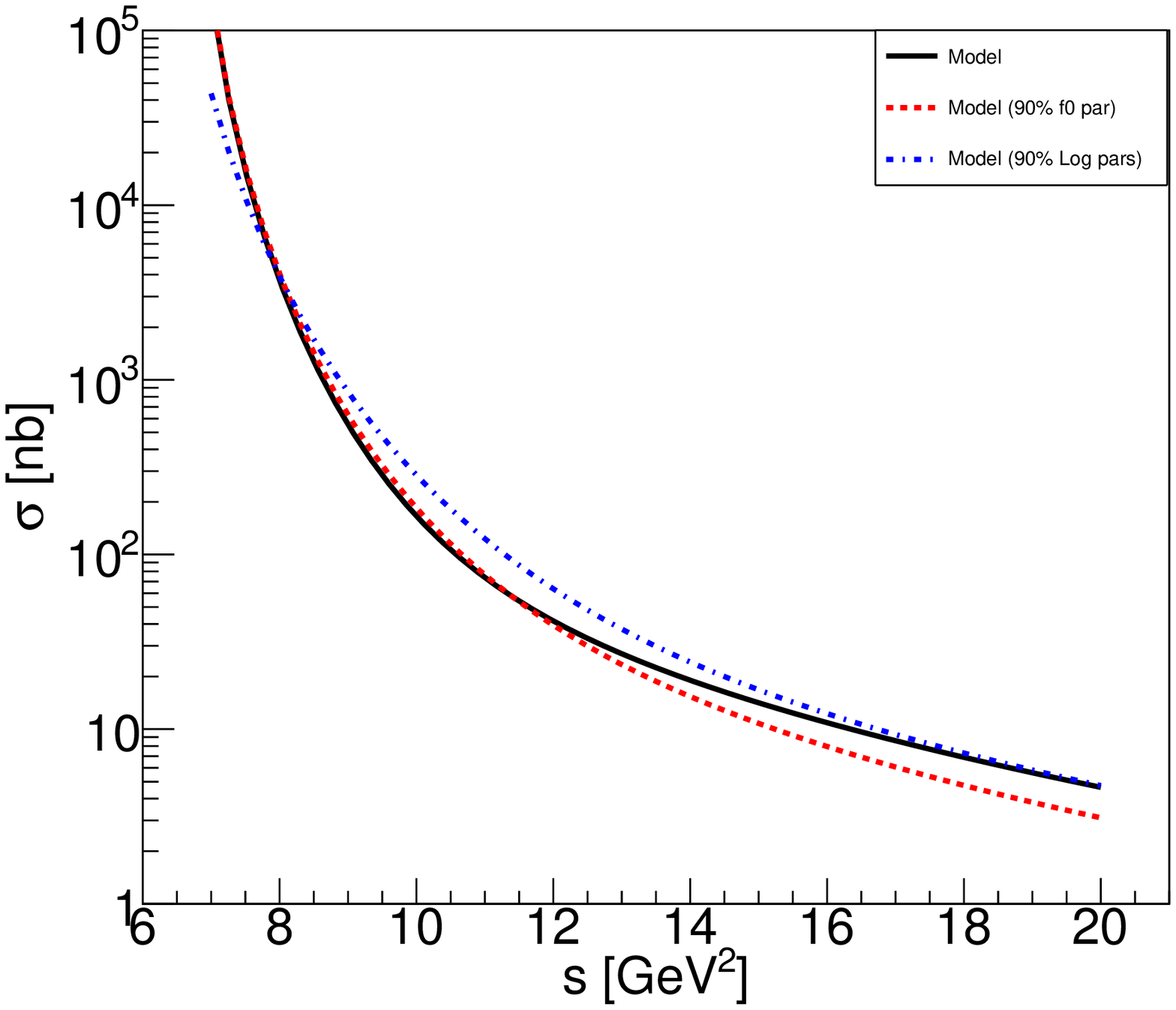}}
\caption{Parameter dependence of the cross section for the reaction $\bar{p} p \rightarrow \pi^0\pi^0$, integrated for 
$|\cos\theta| \le$ 0.66. The calculation with the nominal parameter is shown (black solid line), together with the calculation corresponding to 10\% decrease of the $f_0$ parameters (red dashed line) and  to 10\% decrease of the logarithmic form factor parameters  (blue dash-dotted line).}
\label{Fig:sensitivity}
\end{figure}

The fitted plots and data from Ref. \cite{Armstrong:1997gv} are shown in the Fig. \ref{Fig:lowhigh}, in the energy range 2.911 GeV $\le \sqrt{s}\le$ 3.686 GeV. The data were measured in regular intervals,  with a gap between 3.097 GeV and 3.526 GeV which separate the data into the `lower energy region' (2.911 GeV $\le \sqrt{s}\le$ 3.097 GeV) and `higher energy region'  (3.526 GeV $\le \sqrt{s}\le$ 3.686 GeV). In the lower energy region, a bump produced by higher $L$ resonances appears around $\cos\theta=0$.  It can not be reproduced by the $f_0 $ and $f_2$ mesons considered in $s-$channel, and it disappears at higher energies. We did not attempt to add higher resonances. More precise data are expected from PANDA in a larger angular range, better constraining the model. 
Note that good agreement can be found neglecting the $f_2$ contribution. The $s-$dependence for the cross section of neutral pion production from 5 GeV$^2$ to 20 GeV$^2$ is shown in Fig. \ref{Fig:pi0total-s}, where the experimental point is obtained integrating the data  from Ref. 
\cite{Armstrong:1997gv}  in the available angular range. The calculation is integrated in the same angular range  $0<\cos\theta<  0.66$ or 0.48. The calculation reproduces well the integrated data. Note that the available data cover a reduced angular distribution, whereas the very forward and backward regions give the largest contribution to the total cross section.
In order to appreciate the the sensitivity of the calculation to a selected choice of parameters, in Fig. \ref{Fig:sensitivity} the cross section, integrated for $0<\cos\theta<  0.66$, is reported (black solid line) together with the result of the calculation when decreasing by 10\%  the parameters of $f_0$ (red dashed line) and of the logarithmic form factor (blue dash-dotted line).

%
\subsubsection{The higher energy set}
The case of the set of data at $\sqrt{s}=$4.274 GeV is peculiar. The data correspond to the higher energy available, and  show a discontinuity with respect to the other sets. In particular the bump for $\cos\theta$=0 evolves definitely into a dip. To reproduce this dip, the $L=2$ $f_2$ meson is added. The form factor of $f_2 NN$ is taken as a monopole, Eq. (\ref{eq:f0}), similarly to $f_0$ and the relative phase is also taken as unity.
Concerning  the higher energy, the contribution from $f_0$ meson results  suppressed by the fitting procedure. 
The new parameters for the $s-$channel are listed in the Table \ref{Table:4274}, the other parameters are fixed as in Tables \ref{Table:pich}, and \ref{Table:fixed}.

\begin{figure}[htp!]
\mbox{\includegraphics[width=10.cm]{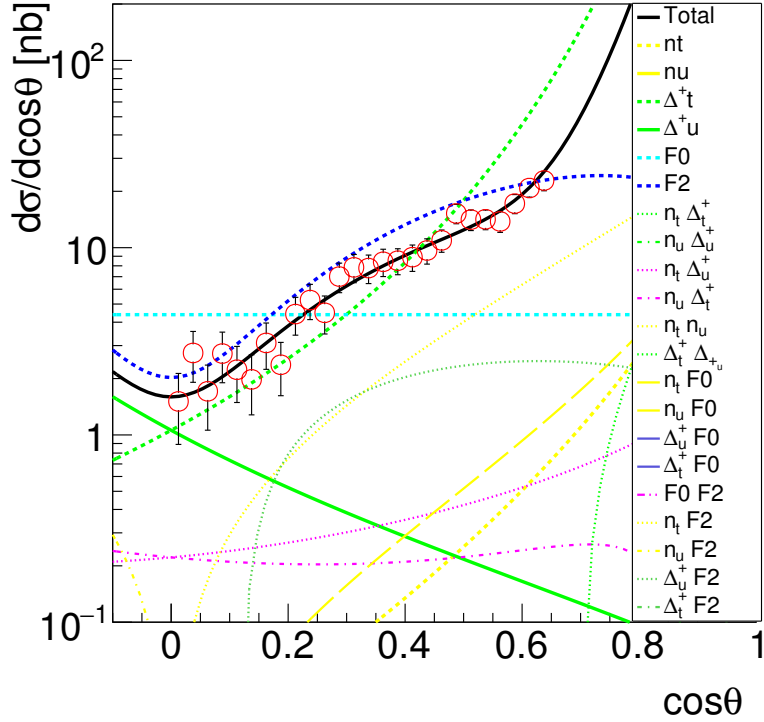}}
\caption{Angular distribution for reaction $\bar p p  \to \pi^0\pi^0$ at  $\sqrt{s}$ = 4.274 GeV \cite{Armstrong:1997gv} with different components. The parameters are listed in Table \ref{Table:4274}.}
\label{Fig:components}
\end{figure}

The different components are visible in Fig. \ref{Fig:components}. One can see that the shape of the angular distribution is very well reproduced by the $f_2$ contribution. The $\Delta$ contribution overcomes the $N$ term. The angular distribution is limited and one can not draw firm conclusions on the $t$ and $u$ channel interplay of the different contributions. A very good agreement is obtained by fitting this set of the data with the present model.

Applying SU(3) symmetry, one can connect other neutral channels. As we see in next section, it works relatively well.
\begin{table}
\begin{tabular}{|c|c|}
\hline
parameters & Value \\
\hline
$F_{f_0}$ & 0.870 $\pm$ 0.014 GeV\\
$F_{f_2}$ & 0.187 $\pm$ 0.001 GeV\\
$\chi^2/ndf$  & 0.787\\
\hline\hline
\end{tabular}
\caption{Parameters of form factors for $f_0$ and $f_2$ mesons at $\sqrt{s}$ = 4.274 GeV.} 
\label{Table:4274}
\end{table}
%
\subsection{The reactions $\bar p+p \to \pi^0 +\eta$ and  $\bar p+p \to \eta +\eta$}

The two body channels :
\ba
\bar p(p_1)+p(p_2)&\to& \eta(k_1) + \eta(k_2),
\label{Eq:eq2} \\
\bar p(p_1)+p(p_2)&\to&  \eta (k_1) + \pi^0(k_2).
\label{Eq:eq3}
\ea
involve mesons that are related by SU(3) symmetry, as $\pi$, $\eta$ and $\eta'$ are members of a single nonet.  Having a model that reproduces consistently angular distributions and cross sections for $\pi^0+\pi^0$,  based on $s$, $t$, and $u$ channels, the amplitudes for the decay to the channels (\ref{Eq:eq1}), (\ref{Eq:eq2}), and (\ref{Eq:eq3})  are related by the SU(3) symmetry. Taking into account that, in principle, $\bar p p$ does not couple directly to $s\bar s$, the following relations hold:
\be
f(\pi^0\eta)= f(\pi^0+\pi^0) \cos\Theta, \  
f(\eta\eta)= f(\pi^0+\pi^0) \cos^2\Theta,
\label{Eq:su3}
\ee
where $\Theta\simeq 45^\circ$ is the pseudoscalar mixing angle  \cite{Singh:2010wd}.

\begin{figure}[htp!]
\mbox{\includegraphics[width=16cm]{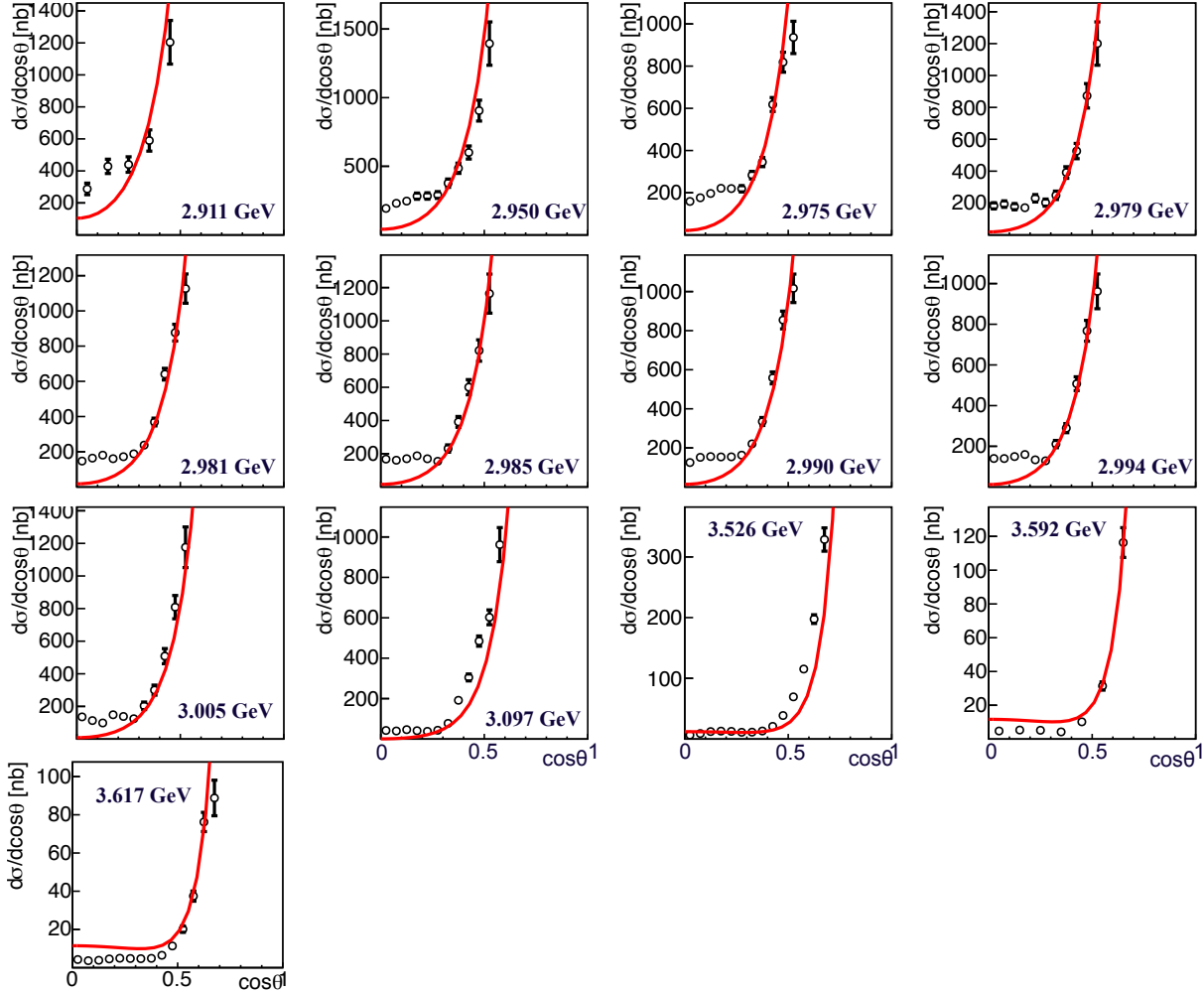}}
\caption{Angular distribution for $\bar{p}p \rightarrow \eta \eta$ (black circle) \cite{Armstrong:1997gv} in the energy range 2.911 GeV$\le \sqrt{s} \le$ 3.617 GeV and the model calculation (red solid curve) based on the symmetry of the quark model.}
\label{Fig:etaeta}
\end{figure}

\begin{figure}[htp!]
\mbox{\includegraphics[width=16cm]{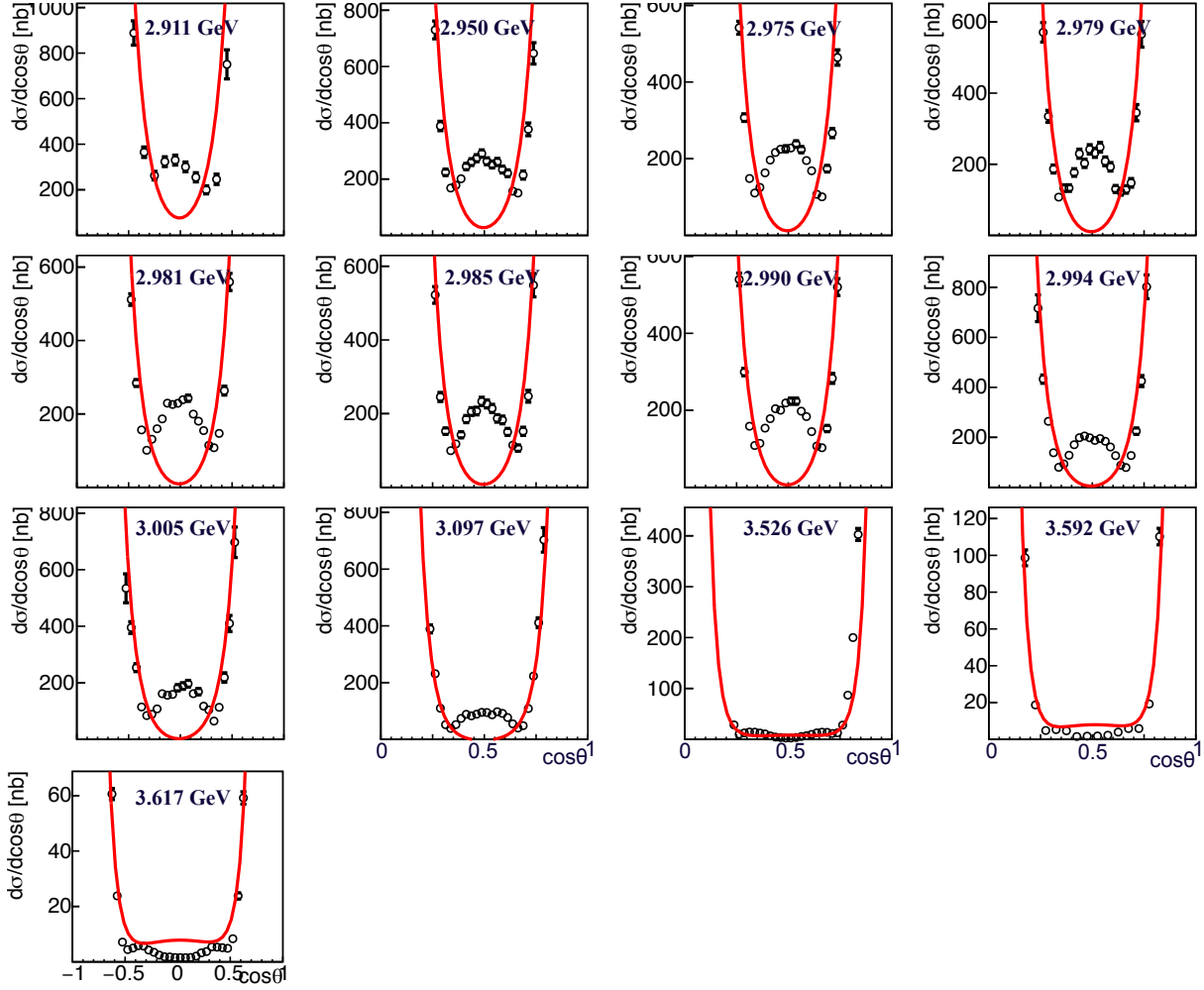}}
\caption{Same as Fig. \ref{Fig:etaeta}, for the reaction $\bar{p}p \rightarrow \eta \pi^0$.}
\label{Fig:etapi0}
\end{figure}

The procedure follows the one derived above for $\pi^0\pi0$. The masses have to be changed correspondingly in the calculation of the kinematics and of the amplitudes. Moreover, in case of reaction (\ref{Eq:eq3}) the fact that the final state is not symmetric induces  a backward-forward asymmetry. Applying SU(3) symmetry and taking into account the kinematics difference due to the masses, the model is applied in the energy range 2.911 GeV$\le \sqrt{s} \le$ 3.617 GeV. The results  are shown  in Fig. \ref{Fig:etaeta} and Fig. \ref{Fig:etapi0}, for the reactions (\ref{Eq:eq2}) and (\ref{Eq:eq3}) respectively. The agreement is very good, without readjusting the parameters. The model is able to reproduce the data in the backward and forward regions.  Similarly to $\pi^0\pi^0$ it is expected that the bump around $\cos\theta=0$ is not described, as it needs to include additional  contributions.


For the higher energy $\sqrt{s}=4.274$ GeV, the data sets for $\eta\eta$ and $\eta \pi^0$ production have large error bars and a few points are measured. Precise data are expected from the PANDA experiment to fill this region.

\section{Conclusions}
A model, built on effective meson Lagrangian, has been applied  to two neutral pion production in proton-antiproton annihilation in the energy range ($2.2 \le \sqrt{s} \le 4.4$) GeV. 

We took a logarithmic form for $s$ and $t(u)$-dependent  form factors.  Coupling constants are fixed from the known properties of the known decay width. The agreement with the existing data  from 
Ref. \cite{Armstrong:1997gv} is satisfactory for the angular dependence as well as the energy dependence of the cross section, especially at high energy. In particular the model is able to describe very nicely the available data for $\pi^0\pi^0$ production 
at $\sqrt{s}=4.274$ GeV.

Around $\cos\theta=0$, the model follows naturally the expected behavior from quark counting rules, concerning the $s$-dependence. However, the bump in the central region, present at low energies, is missed by the model. Possible improvement is foreseen by adding other components, that however, should vanish as the energy increases. A 'fine tuning' is desirable, and will be more meaningful  when more data will be available at PANDA, in a larger and more complete angular and energy range.  The implementation to Monte Carlo simulations for predictions and optimization to the forecoming PANDA experiment is foreseen for this aim, too.
 
Using SU(3) symmetry, without any change of parameters, the angular distributions for $\bar p+p \to \eta +\eta$, and  for the asymmetric reaction $\bar p+p \to \pi^0 +\eta$ are recovered.  

\section{Acknowledgments}
Thanks are due to D. Marchand and A.E. Dorokhov, for useful discussions and interest in this work. One of us (Yu.B) acknowledges kind hospitality at IPN Orsay, in frame of JINR-IN2P3 agreement and the support of the Heisenberg-Landau program (HLP-2017-11).

\section{Appendix}
The relevant formulas for the amplitudes and their interferences are given below.
\subsection{$s$-exchange of neutral scalar mesons: the $f_0$ contribution}
The matrix element is written as: 
\be
i{\cal M}_{f_0}=
  -\displaystyle\frac{g_{f_0 N N }g_{f_0  \pi\pi}}
 {q_s^{2}-m_{f_0}^{2}+i \sqrt{q_s^2}\Gamma_{f_0}(q_s^2) }
 \bar v (p_1) u(p_2).
 \label{Eq:Mf0}
 \ee
 Squaring the amplitude one finds:
\be
 |{\cal M}|^2 
 =\displaystyle\frac{g_{f_0 N N }^2g_{f_0  \pi\pi}^2 }
 {|q_s^{2}-m_{f_0}^{2}+i \sqrt{q_s^2}\Gamma_{f_0}(q_s^2) |^2} 2(s-4M^2).
  \label{Eq:Mf02}
 \ee

\subsubsection{The $f_0 \pi \pi$ coupling constant}
 
The decay width of the $f_0$ meson in the system where it is at rest is given by :
\be
d\Gamma(f_0\to \pi\pi)  = 
 \displaystyle\frac{1}{2m_{f_0}} | {\cal M}(f_0\to \pi\pi)|^{2} d\Phi_2 ,\ 
 \label{Eq:gam}
 \ee
 with the phase space:
 \be
d \Phi_2=\displaystyle\frac{\Lambda^{1/2}(m_{f_0},m_\pi,m_\pi)}{2^5\pi^2 m_{f_0}^2} d \Omega,\ \Lambda^{1/2}(m_{f_0},m_\pi,m_\pi)= M^2\sqrt{1- \displaystyle\frac{4 m_{\pi}^2}{M_p^2} }.
\ee
Therefore: 
\be
\Phi_2=\displaystyle\frac{\Lambda^{1/2}(m_{f_0},m_\pi,m_\pi)}{2^3\pi} \sqrt{1- \displaystyle\frac{4 m_{\pi}^2}{M_p^2} }.
 \ee

The matrix element  for the decay $f_0\to \pi \pi$ is (see Fig. \ref{Fig:f02vertex}):
\be
 {\cal M}(f_0\to \pi \pi)=\displaystyle\frac{1}{(2\pi)^4}  g_{f_0 \pi \pi}.
\label{Eq:am}
\ee

\subsubsection{The  $f_0$ interferences}
\begin{enumerate}
\item{The $N-f_0 $ interference}
\be
2Re[{\cal M}_N^*{\cal M}_{f_0}]= 
2Re \
\displaystyle\frac{ g_{f_0 NN } g_{f_0 \pi\pi} g^2_{\pi NN}} 
{[ s-m_{f_0}^2-i \sqrt{s} \Gamma_{f_0}(s) ] (t-M_p^2)} 
Tr \left  [ (\hat p_1 - M_p) (-\hat q_t+M_p)   (\hat p_2 + M_p) 
\right ].
\label{Eq:intNf0}
\ee 
\item{The $\Delta - f_0 $ interference }
\ba
2Re[{\cal M}_{\Delta}^*{\cal M}_{f_0}]&=& 
2Re \ 
\displaystyle\frac{ g_{f_0 NN } g_{f_0 \pi \pi} g^2_{\Delta N N}} 
{[ s-m_{f_0}^2-i \sqrt{s}\Gamma_{f_0}(s) ] (t-M_{\Delta}^2)} \nn\\
&&Tr \left  [ (\hat p_1 - M_p)  (-\hat q_t+M_{\Delta})  \tilde P_{\alpha \beta}  (\hat p_2 + M_p) \right ]
k_1^\alpha k_2^\beta.
\label{Eq:intDf0}
\ea
\end{enumerate}
\subsection{$s$-exchange of neutral scalar mesons: the $f_2$ contribution}
Let us consider $f_2(1270)$ with spin 2 and positive parity, that decays $\sim 100\%$  into two neutral pions.
\subsubsection{The $f_2$- propagator} 

 The $f_2$- propagator is 
 \be
\displaystyle\frac{ \chi_{\mu\nu} \chi_{\alpha\beta}}{q_s^{2}-m_{f_2}^{2}+i \sqrt{q_s^2}\Gamma_{f_2}(q_s^2) },
\label{Eq:am}
\ee
where the width of the $f_2$ is taken into account by the Breit-Wigner function and the transferred momentum is $q_s=p_1+p_2=k_1+k_2$, $q_s^2=s$.
\subsubsection{The vertex $f_2\to p p$ }
The Lagrangian for the decay  $f_2\to p p$ is written as: 
\be
{\cal L}_{f_{2}\bar pp}=g_{f_{2}pp}\bar p (\gamma_\mu  i \partial_\nu  + \gamma_\nu  i \partial _\mu  
+ \displaystyle\frac{2}{3} \eta _{\mu\nu}  i\hat \partial   ) p T^{\mu\nu}.
 \label{Eq:L1}
 \ee
 The last term in Eq.  (\ref{Eq:L1}) vanishes as it is the product of an antisymmetric and a symmetric tensor.
 \begin{enumerate}
 \item The vertex $f_2\to p p$ then is written as (see Fig. \ref{Fig:DiaAll}c)
\be
(-i) g_{f_{2}pp}\gamma_\mu (p_1-p_2)_\nu \chi^{\mu\nu}, 
 \label{Eq:gam}
 \ee
where the symmetric tensor $\chi^{\mu\nu}$ has the following properties: 
\be
 \chi_{\mu\nu} =\chi_{\nu\mu},\  \chi_{\mu\nu} g^{\mu\nu}=0, \  \chi_{\mu\nu}q^\nu =0;\  
 \chi_{\mu\nu} \chi_{\alpha\beta}=  \displaystyle\frac{1}{2}\left (  \eta _{\alpha\nu}  \eta_{\nu\beta} +
  \eta _{\mu\beta}  \eta_{\nu\alpha}  \right )-  \displaystyle\frac{1}{3} \eta_{\mu\nu} \eta_{\alpha\beta},
 \label{Eq:chi}
 \ee
with  $\eta _{\mu\nu} =- g _{\mu\nu} + \displaystyle\frac{q_\mu q_\nu}{q^2}$, and $q$  is the $f_2$ meson four momentum.
\end{enumerate}

\subsubsection{The vertex $f_{2 \pi \pi}$ }

The amplitude for the  $f_2\to \pi\pi$ decay is:
\be
 {\cal M} (f_2 \to \pi \pi)=-\displaystyle\frac{-1}{(2\pi)^4} (-i) g_{f_2 \pi \pi} \chi^{\mu\nu}\Theta ^\pi_{\mu\nu},
\label{Eq:am1}
\ee
where $ g_{f_2  \pi\pi}$ is the constant for the decay $f_2 \to  \pi\pi$ and 
\be
 \Theta ^\pi_{\mu\nu}=\displaystyle\frac{1}{2}   \eta_{\mu\nu} (\partial_\alpha\pi)^2- (\partial_\mu\pi )(\partial_\nu\pi ),
\label{Eq:am2}
\ee
which results in:
\ba
 {\cal M} (f_2 \to \pi \pi)&=&(-i) \displaystyle\frac{1}{2} g_{f_2 \pi \pi} \chi^{\mu\nu} \left [2 \displaystyle\frac{1}{2}\eta_{\mu\nu} (k_1k_2) - k_{1\mu}k_{2\nu}-k_{1\nu}k_{2\mu}\right ]=\nn\\
&&\displaystyle\frac{i}{2}g_{f_2 \pi \pi} \chi^{\mu\nu} \left [ k_{1\mu}k_{2\nu}+k_{1\nu}k_{2\mu} - (k_1k_2) \eta_{\mu\nu}\right ].
\label{Eq:am3}
\ea
The matrix element for $f_2$ $s-$channel exchange in $\bar pp\to \pi^0\pi^0$ is :
\ba
{\cal M}_{f_{2}}&=&\displaystyle\frac{g_ {f_{2}pp} g_{f_{2} \pi \pi} }{2}
 \left [\bar v(p_1) \gamma_\mu (p_2-p_1)_\nu u(p_2) \right ]\times
\nn\\
&&
\displaystyle\frac{F^{\mu\nu\alpha\beta} }{s-m_{f_{2}}^{2}+i \sqrt{s}\Gamma_{f_2}}
 [k_{1\alpha} k_{2\beta}+ k_{1\beta}k_{2\alpha}-(k_1k_2)\eta_{\alpha\beta}],\nn\\
 {\cal M}_{f_{2}}^*&=&
\displaystyle\frac{g_{f_{2}pp} g_{f_{2} \pi \pi}} {2}
\left [\bar u(p_2) \gamma_\rho (p_2-p_1)_\sigma v(p_1)  \right )]\times
\nn\\
&&
\displaystyle\frac{F^{\rho\sigma \gamma\delta }}{s-m_{f_2}^{2}-i \sqrt{s}\Gamma_{f_2}}
 [k_{1\gamma} k_{2\delta}+ k_{1\delta}k_{2\gamma}-(k_1k_2)\eta_{\gamma\delta}],\nn
\ea
where $F^{\mu\nu\alpha\beta}= \chi_{\mu\nu} \chi_{\alpha\beta}$. 
The matrix element squared is:
\ba
|{\cal M}_{f_{2}}|^2&= 
&\displaystyle\frac{g_ {f_{2}pp}^2 g_{f_2 \pi \pi}^2} {4}
\displaystyle\frac{F^{\mu\nu\alpha\beta} F^{\rho\sigma \gamma\delta}}
{|s-m_{f_2}^{2}+i \sqrt{s}\Gamma_{f_2}|^2}\times
\nn\\
&&
Tr [(\hat p_1-M) \gamma_\mu (p_2-p_1)_\nu (\hat p_2+M)\gamma_\rho (p_2-p_1)_\sigma]\times
\nn\\
&&
[k_{1\alpha } k_{2\beta }+ k_{1\beta } k_{2\alpha }-(k_1k_2) \eta_{\alpha\beta} ]
   [k_{1\gamma} k_{2\delta}+ k_{1\delta}k_{2\gamma}-(k_1k_2)\eta_{\gamma\delta}].
\label{Eq:am5}
\ea

The decay width of the $f_2$ meson in the system where it is at rest is given by :
\be
d\Gamma(f_2\to \pi\pi)  = 
 \displaystyle\frac{g_{f_2 \pi \pi}^2}{2m_{f_2}} | {\cal M}(f_2\to \pi\pi)|^{2} d\Phi_2 ,\ 
 \label{Eq:gam}
 \ee
 with the phase space:
 \be
d \Phi_2=\displaystyle\frac{\Lambda^{1/2}(m_{f_{2}},m_\pi,m_\pi)}{2^5\pi^2 m_{f_{2}}^2} d \Omega,\ \Lambda^{1/2}(m_{f_{2}},m_\pi,m_\pi)= m_{f_{2}}^2\sqrt{1- \displaystyle\frac{4 m_{\pi}^2}{m_{f_{2}}^2} }.
 \label{Eq:psf2}
 \ee
Therefore: 
\be
\Phi_2=
\displaystyle\frac{\Lambda^{1/2}(m_{f_{2}},m_\pi,m_\pi)}{2^3\pi} 
\sqrt{1- \displaystyle\frac{4 m_{\pi}^2}{m^2_{f_{2}}}}.
 \label{Eq:phi2}
 \ee
 
The matrix element  for the decay $f_2\to \pi \pi$ is ( see Fig. \ref{Fig:f02vertex}c)
\ba
|{\cal M}(f_{2}\to \pi \pi)|^2&= &
\displaystyle\frac{1} {4}
\displaystyle\frac{F^{\mu\nu\alpha\beta} F^{\rho\sigma \gamma\delta }}
{|s-m_{f_2}^{2}+i \sqrt{s}\Gamma_{f_2}|^2}
[k_{1\alpha } k_{2\beta }+ k_{1\beta } k_{2\alpha }-(k_1k_2) \eta_{\alpha\beta } ]\times \nn\\
 &&[k_{1\gamma} k_{2\delta}+ k_{1\delta}k_{2\gamma}-(k_1k_2)\eta_{\gamma\delta}].
\label{Eq:am6}
\ea
\be
 \Gamma_{f_{2}}\ = \displaystyle\frac{g_{f_2 \pi \pi}^2}{16m_{f_2}\pi} |{\cal M}(f_{2}\to \pi \pi)|^2\sqrt{1- \displaystyle\frac{4 m_{\pi}^2}{m_{f_{2}}^2} }.
\label{Eq:phaseR}   
\ee
Taking the value: $\Gamma= (0.1867\pm 0.0025) $ GeV, one finds  $g_{f_2\pi\pi}= (19 \pm 0.26) $ GeV$^{-1}$. 
\subsubsection{The $f_2$ interferences}
\begin{enumerate}
\item
 The $f_0-f_2 $ interference 
\ba
2Re[ {\cal M}^*_{f_{0}} {\cal M}_{f_{2}}]&=& 
-Re 
\displaystyle\frac{ g_{f_{0} NN } g_{f_{0} \pi\pi} g_{f_{2} NN } g_{f_{2} \pi\pi} } 
{[ s-m_{f_{0}}^2-i \sqrt{s} \Gamma_{f_{0}}(s) ] [ s-m_{f_{2}}^2+i \sqrt{s} \Gamma_{f_{2}}(s) ] } 
F^{\mu\nu\alpha\beta}\label{Eq:intNf0}\\
&&[k_{1\alpha} k_{2\beta}+ k_{1\beta}k_{2\alpha}-(k_1k_2)\eta_{\alpha\beta}]Tr \left  [ (\hat p_1 - M_p) \gamma_\mu(p_2-p_1)_\nu   (\hat p_2 + M_p) 
\right ] .
\nn
\ea
\item{The $N-f_2 $ interference }
\ba
2Re[ {\cal M}_{f_{2}} {\cal M}_N^*]&=& 
Re 
\displaystyle\frac{ g_{f_{2} NN } g_{f_{2} \pi\pi} g^2_{\pi NN}} 
{[ s-m_{f_{2}}^2+i \sqrt{s} \Gamma_{f_{2}}(s) ] (t-M_p^2)} 
F^{\mu\nu\alpha\beta}[k_{1\alpha} k_{2\beta}+ k_{1\beta}k_{2\alpha}-\nn\\
&& (k_1k_2)\eta_{\alpha\beta}]Tr \left  [ (\hat p_1 - M_p) \gamma_\mu(p_2-p_1)_\nu   (\hat p_2 + M_p)  (-\hat q_t+M_p)   \right ].
\label{Eq:intNf0}
\ea
\item{The $\Delta - f_2 $ interference }
\ba
2Re[ {\cal M}_{\Delta}^* {\cal M}_{f_{2}} ]&=& 
-Re
\displaystyle\frac{ g_{f_{2} NN } g_{f_{2} \pi \pi} g^2_{N\Delta N \pi}} 
{[ s-m_{f_{2}}^2+i \sqrt{s}\Gamma_{f_{2}}(s) ] (t-M_{\Delta}^2)} 
\nn\\
&& F^{\mu\nu\alpha\beta}(k_{1\alpha} k_{2\beta}+ k_{1\beta}k_{2\alpha}-(k_1k_2)\eta_{\alpha\beta})\nn\\
&&Tr \left  [ (\hat p_1 - M_p)  \gamma_\mu(p_2-p_1)_\nu (\hat p_2 + M_p) P_{\rho\sigma}(q) (\hat q_t+M_{\Delta})   \right ]
k_1^\sigma k_2^\delta .
\label{Eq:intDf0}
\ea
\end{enumerate}

\begin{thebibliography}{11}
\expandafter\ifx\csname natexlab\endcsname\relax\def\natexlab#1{#1}\fi
\expandafter\ifx\csname bibnamefont\endcsname\relax
  \def\bibnamefont#1{#1}\fi
\expandafter\ifx\csname bibfnamefont\endcsname\relax
  \def\bibfnamefont#1{#1}\fi
\expandafter\ifx\csname citenamefont\endcsname\relax
  \def\citenamefont#1{#1}\fi
\expandafter\ifx\csname url\endcsname\relax
  \def\url#1{\texttt{#1}}\fi
\expandafter\ifx\csname urlprefix\endcsname\relax\def\urlprefix{URL }\fi
\providecommand{\bibinfo}[2]{#2}
\providecommand{\eprint}[2][]{\url{#2}}

\bibitem[{\citenamefont{Wang et~al.}(2017)\citenamefont{Wang, Bystritskiy, and
  Tomasi-Gustafsson}}]{Wang:2015ybw}
\bibinfo{author}{\bibfnamefont{Y.}~\bibnamefont{Wang}},
  \bibinfo{author}{\bibfnamefont{Y.~M.} \bibnamefont{Bystritskiy}},
  \bibnamefont{and}
  \bibinfo{author}{\bibfnamefont{E.}~\bibnamefont{Tomasi-Gustafsson}},
  \bibinfo{journal}{Phys. Rev.} \textbf{\bibinfo{volume}{C95}},
  \bibinfo{pages}{045202} (\bibinfo{year}{2017}).

\bibitem[{\citenamefont{Lutz et~al.}(2009)}]{Lutz:2009ff}
\bibinfo{author}{\bibfnamefont{M.}~\bibnamefont{Lutz}} \bibnamefont{et~al.}
  (\bibinfo{collaboration}{PANDA Collaboration}) (\bibinfo{year}{2009}),
  \eprint{0903.3905 [hep-ph]}.

\bibitem[{\citenamefont{Loiseau}(1992)}]{Loiseau:1992xe}
\bibinfo{author}{\bibfnamefont{B.}~\bibnamefont{Loiseau}},
  \bibinfo{journal}{Nucl. Phys.} \textbf{\bibinfo{volume}{A543}},
  \bibinfo{pages}{33C} (\bibinfo{year}{1992}).

\bibitem[{\citenamefont{Olive et~al.}(2014)}]{Agashe:2014kda}
\bibinfo{author}{\bibfnamefont{K.}~\bibnamefont{Olive}} \bibnamefont{et~al.}
  (\bibinfo{collaboration}{Particle Data Group}), \bibinfo{journal}{Chin.Phys.}
  \textbf{\bibinfo{volume}{C38}}, \bibinfo{pages}{090001}
  (\bibinfo{year}{2014}).

\bibitem[{\citenamefont{Armstrong et~al.}(1997)}]{Armstrong:1997gv}
\bibinfo{author}{\bibfnamefont{T.}~\bibnamefont{Armstrong}}
  \bibnamefont{et~al.} (\bibinfo{collaboration}{Fermilab E760 Collaboration}),
  \bibinfo{journal}{Phys. Rev.} \textbf{\bibinfo{volume}{D56}},
  \bibinfo{pages}{2509} (\bibinfo{year}{1997}).

\bibitem[{\citenamefont{Machleidt}(2001)}]{PhysRevC.63.024001}
\bibinfo{author}{\bibfnamefont{R.}~\bibnamefont{Machleidt}},
  \bibinfo{journal}{Phys. Rev. C} \textbf{\bibinfo{volume}{63}},
  \bibinfo{pages}{024001} (\bibinfo{year}{2001}).

\bibitem[{\citenamefont{Chernyak and
  Zhitniskii}(1977)}]{chernyak1977asymptotic}
\bibinfo{author}{\bibfnamefont{V.~L.} \bibnamefont{Chernyak}} \bibnamefont{and}
  \bibinfo{author}{\bibfnamefont{A.~R.} \bibnamefont{Zhitniskii}},
  \bibinfo{journal}{JETP Lett} \textbf{\bibinfo{volume}{25}}
  (\bibinfo{year}{1977}).

\bibitem[{\citenamefont{Lepage and Brodsky}(1979)}]{lepage1979exclusive}
\bibinfo{author}{\bibfnamefont{G.~P.} \bibnamefont{Lepage}} \bibnamefont{and}
  \bibinfo{author}{\bibfnamefont{S.~J.} \bibnamefont{Brodsky}},
  \bibinfo{journal}{Phys. Rev.Lett.} \textbf{\bibinfo{volume}{43}},
  \bibinfo{pages}{545} (\bibinfo{year}{1979}).

\bibitem[{\citenamefont{Ablikim et~al.}(2014)}]{Ablikim:2014jrz}
\bibinfo{author}{\bibfnamefont{M.}~\bibnamefont{Ablikim}} \bibnamefont{et~al.}
  (\bibinfo{collaboration}{BESIII Collaboration}), \bibinfo{journal}{Phys.
  Lett.} \textbf{\bibinfo{volume}{B735}}, \bibinfo{pages}{101}
  (\bibinfo{year}{2014}).

\bibitem[{\citenamefont{Aubert}(2006)}]{PhysRevD.73.012005}
\bibinfo{author}{\bibfnamefont{B.}~\bibnamefont{Aubert}}
  (\bibinfo{collaboration}{BABAR Collaboration}), \bibinfo{journal}{Phys.
  Rev.D} \textbf{\bibinfo{volume}{73}}, \bibinfo{pages}{012005}
  (\bibinfo{year}{2006}).

\bibitem[{\citenamefont{Singh et~al.}(2011)\citenamefont{Singh, Lee, and
  Wang}}]{Singh:2010wd}
\bibinfo{author}{\bibfnamefont{J.~P.} \bibnamefont{Singh}},
  \bibinfo{author}{\bibfnamefont{F.~X.} \bibnamefont{Lee}}, \bibnamefont{and}
  \bibinfo{author}{\bibfnamefont{L.}~\bibnamefont{Wang}},
  \bibinfo{journal}{Int. J. Mod. Phys.} \textbf{\bibinfo{volume}{A26}},
  \bibinfo{pages}{947} (\bibinfo{year}{2011}).

\end{thebibliography}

\end{document}